\documentclass{osa-article}
\journal{oe}
\articletype{Research Article}

\begin{document}

\title{Electronic Phase Detection with sub-10~fs Timing Jitter for Terahertz Time-Domain Spectroscopy Systems}

\author{Felix Paries,\authormark{1,2} Oliver Boidol,\authormark{1} Georg von Freymann,\authormark{1,2} and Daniel Molter\authormark{1,*}}

\address{\authormark{1}Fraunhofer Institute for Industrial Mathematics ITWM, Fraunhofer-Platz 1, 67663 Kaiserslautern, GERMANY\\
\authormark{2}Department of Physics and Research Center OPTIMAS, Technische Universit\"at Kaiserslautern, 67663 Kaiserslautern, GERMANY}

\email{\authormark{*}daniel.molter@itwm.fraunhofer.de} 


\begin{abstract}

Terahertz time-domain spectroscopy systems based on resonator-internal repetition-rate modulation, such as SLAPCOPS~\cite{Kolano2018} and ECOPS~\cite{Yahyapour2019}, rely on electronic phase detectors which are typically prone to exhibit both a non-negligible random and systematic timing error. This limits the quality of the recorded information significantly. Here, we present the results of our recent attempt to reduce these errors in our own electronic phase detection systems. A more than six-fold timing-jitter reduction from 59.0~fs to 8.6~fs led to a significant increase in both exploitable terahertz bandwidth and signal-to-noise ratio. Additionally, utilizing our interferometrically monitored delay line as a calibration standard, the systematic error could be removed almost entirely and thus, excellent resolution of spectral absorption lines be accomplished. These improvements increased the accuracy of our multi-layer thickness measurements based on electronic phase detection by more than a factor of five, pushing the overall performance well into the sub-µm regime.
\end{abstract}


\section{Introduction}
Over the last decade, terahertz time-domain spectroscopy~\cite{NeuSchmuttenmaer2018} has evolved into an established technology used for a variety of industrial applications such as multi-layer thickness measurements of automotive coatings~\cite{Ellrich2020, Krimi2016, Pfeiffer2018}, monitoring of drying processes~\cite{Klier2021}, or mail inspection~\cite{Molter2021}. The idea is to resolve a sample's terahertz-pulse response in the time domain making both amplitude and phase information accessible which in turn presents the opportunity to calculate the optical path length and the absorption. Since the desired information is only available on a sub-picosecond timescale, a sampling approach based on two ultrashort laser pulse trains is required: one pulse train is used to generate a train of terahertz pulses by repeatedly short-circuiting a semiconductor-based photoconductive switch to which a high voltage is applied~\cite{Auston1984, Fattinger1989, Kohlhaas2019}. Simultaneously, a second laser pulse train is used to sample the electric field of the generated terahertz pulse train by repeatedly short-circuiting a second photoconductive switch while measuring the induced electrical current. For this pump-probe-approach to work, the phase difference between the two pulse trains must be measured very precisely. Traditionally, this has been realized through the use of external delay lines that can be controlled and monitored with sub-2~fs accuracy via interferometric measurement techniques~\cite{Molter2017}. However, due to the inertia of the delay-line's mechanical components, the scan rates are usually limited to below 50 Hz~\cite{Molter2017}. Therefore, these approaches are not able to cover applications where experimental conditions alter on a millisecond time scale. To access this time regime, SLAPCOPS~\cite{Kolano2018} and ECOPS~\cite{Yahyapour2019} systems utilizing resonator-internal repetition-rate modulation have been developed. These delay-line-free systems can achieve scan rates up to 1600 terahertz pulse traces per second~\cite{Yahyapour2019} but come with the disadvantage of having to measure the phase difference electronically via fast photo detection which is prone to exhibit both a non-negligible random and systematic timing error. This in turn yields a quality decrease in the recorded terahertz spectra. For this reason, an improvement of existing electronic phase detection systems is needed. 

In this paper, we present our recent attempt to reduce random and systematic errors in our own electronic phase detection system. As will be seen, the reduction of random errors led to a significant increase in both exploitable terahertz bandwidth and signal-to-noise ratio. Furthermore, by using an interferometry-based phase-difference detection as a measurement standard, we could calibrate our system and thus, remove almost any systematic error. Consequently, the drawbacks of our electronic phase detection unit could be reduced resulting in the recording of terahertz spectra that can compete with those recorded by the use of our external delay lines. To demonstrate the impact on practical applications, we show that these improvements yield a more than five-fold accuracy enhancement of our multi-layer thickness measurements based on electronic phase detection. 

\section{Experimental setup}
\begin{figure}[h!]
    \centering\includegraphics[width=0.9\textwidth]{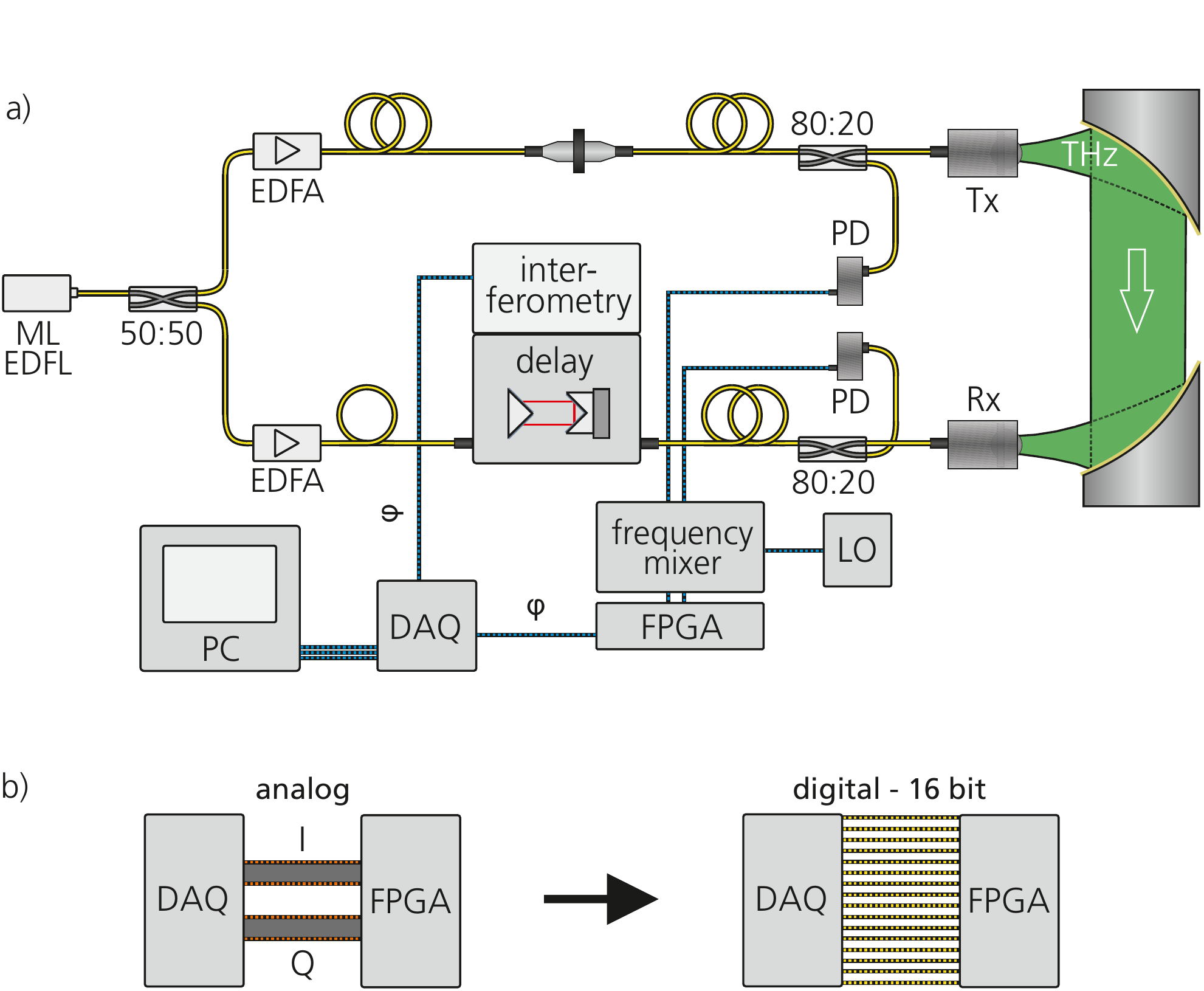}
    \caption{a) Experimental setup: Using a fiber-coupled splitter (50:50), the output of an mode-locked Erbium-doped fiber laser (ML EDFL) is split into two equal parts which are amplified by an Erbium-doped fiber amplifier (EDFA), respectively. One pulse train is guided to a photoconductive terahertz transmitter (Tx), whereas the other one passes a mechanical delay line before being guided to a photoconductive terahertz receiver (Rx). 20\% of each pulse train is outcoupled for photo detection (PD). The phase difference is measured twice: via an interferometric and via an electronic approach. This allows for a comparison between the interferometric phase detection (measurement standard) and the electronic phase detection. b) Previous analog phase transmission architecture vs. new 16-bit digital phase transmission architecture. Switching from an analog to a digital phase transmission eliminated the digital-to-analog-conversion jitter and opened doors for a bandwidth-controlled jitter reduction.}
    \label{fig:setup}
\end{figure}

During the experiments, we utilized the terahertz time-domain spectroscopy setup schematically shown in Fig.~\ref{fig:setup}. To exclude errors caused by the laser-intrinsic timing jitter of a dual-laser setup, we used the output of a single mode-locked ultrashort-pulsed Erbium-doped fiber laser with a repetition rate of 100~MHz (ELMO, Menlo Systems GmbH) and split it into two equal pulse trains via a standard fiber-coupled 50:50 splitter. Each pulse train is then amplified by an Erbium-doped fiber amplifier (ELMA, Menlo Systems GmbH). One pulse train is guided to a photoconductive terahertz transmitter, whereas the other one passes an interferometrically monitored delay line before being guided to a photoconductive terahertz receiver~\cite{Weber2021}. Additionally, two fiber-coupled 80:20 splitters are used to send 20\% of each branch to a separate photodetector, respectively. This setup allows for a twofold measurement of the pulse trains' phase difference: via an interferometric approach determining the delay-line position with a precision of 1.1~fs~\cite{Molter2017}, and via an electronic approach utilizing a fast photo detection of the pulse trains' higher-harmonics' difference signal in combination with a frequency mixer. In this way, the accurate interferometric phase detection can serve as a measurement standard allowing to compare the electronic phase detection against. This enabled us to examine both the random and systematic errors of the electronic phase detection as well as their implications on the resulting terahertz spectra. As a consequence of these experiments, we adjusted our electronic phase detection unit from a previously used analog I-Q phase transmission to a new 16-bit digital phase transmission. This paved the way for a significant reduction of random and systematic timing error which in turn causes a significant terahertz performance enhancement. 

\newpage
\section{Simulation}

\begin{figure}[h!]
    \centering\includegraphics[width=0.9\textwidth]{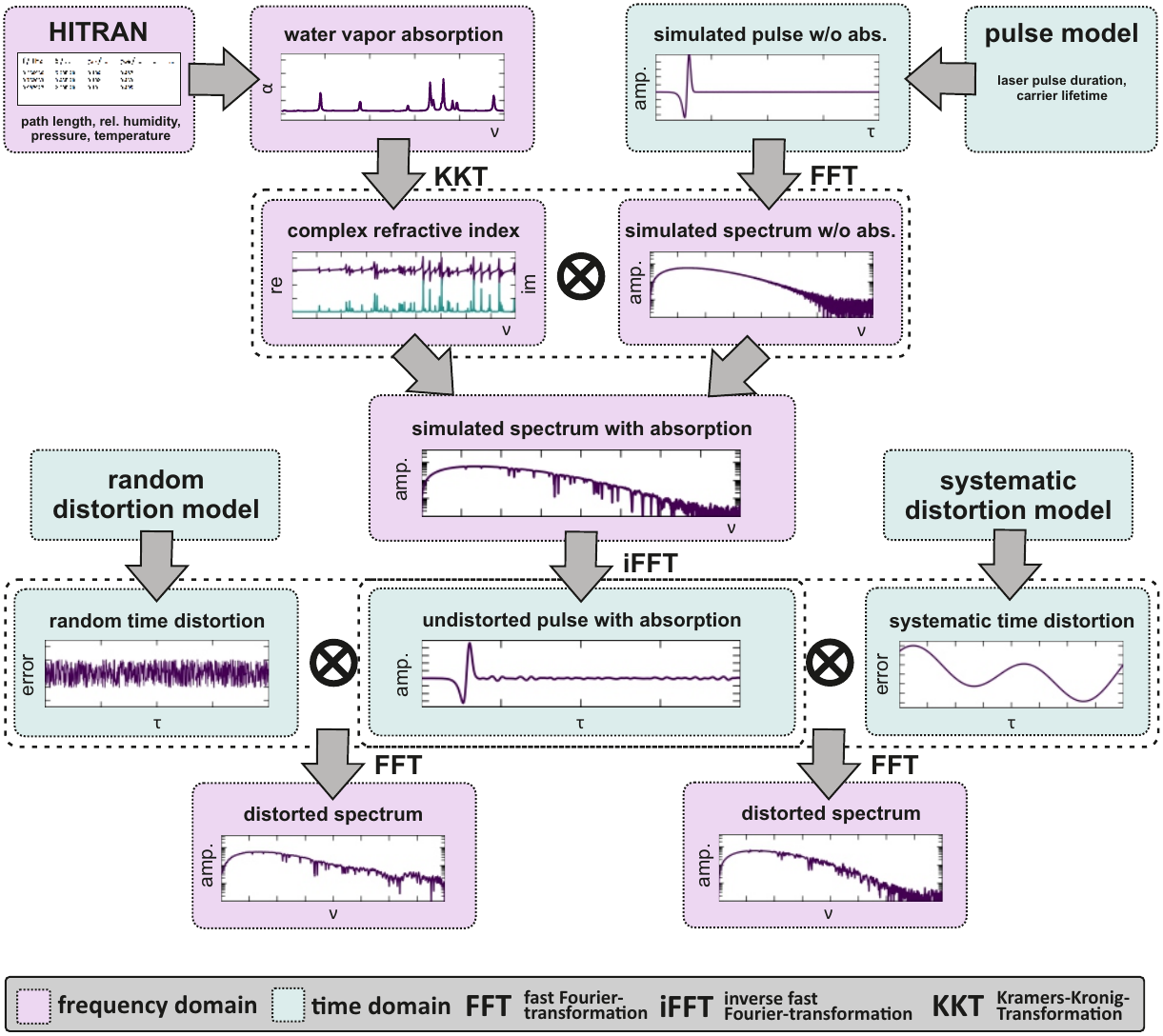}
    \caption{Design of the conducted simulation. Water vapor absorption information from the HITRAN database is imprinted onto a simulated terahertz spectrum and back-transformed into the time domain. Dependent on the model, either a random or a systematic distortion is applied to the time axis of the terahertz pulse containing the absorption information. A final Fourier transformation reveals the influence on the terahertz spectrum.}
    \label{fig:designSimulation}
\end{figure}

Random and systematic errors influence the quality of the recorded terahertz spectra in different ways. In order to substantiate our understanding of these influences, we conducted various numerical simulations. The simulation's design is shown in Fig.~\ref{fig:designSimulation}. Starting with data from the well-known HITRAN database~\cite{HITRAN}, we calculated the terahertz absorption spectra for moist air under consideration of the terahertz optical path length, the relative humidity, the air pressure, and the temperature. The Kramers-Kronig-relation~\cite{KKR} is then used to transform the absorption coefficient into a complex refractive index. Additionally, the transformation of an ultra-short laser pulse to an ideal terahertz pulse is modelled by taking into account the laser pulse duration and the carrier lifetime in the semiconductor of a photoconductive switch. The Fourier transformation of the derived terahertz pulse then yields an absorption-free terahertz spectrum. To impose the water vapor's absorption characteristics onto the absorption-free terahertz spectrum, a convolution of the complex refractive index and the terahertz spectrum is applied. This yields a simulated absorption spectrum which can be back-transformed into the time domain via an inverse Fourier transformation to obtain a terahertz pulse containing the absorption information. By applying a distortion model to the time axis, it is now possible to simulate an erroneous phase detection and examine the influence of both a time-axis' random and static error. As a last step, the newly created distorted pulse is Fourier-transformed to obtain the desired distorted terahertz absorption spectrum. 

\subsection{Simulation of random error}

\begin{figure}[h!]
    \centering\includegraphics[width=0.99\textwidth]{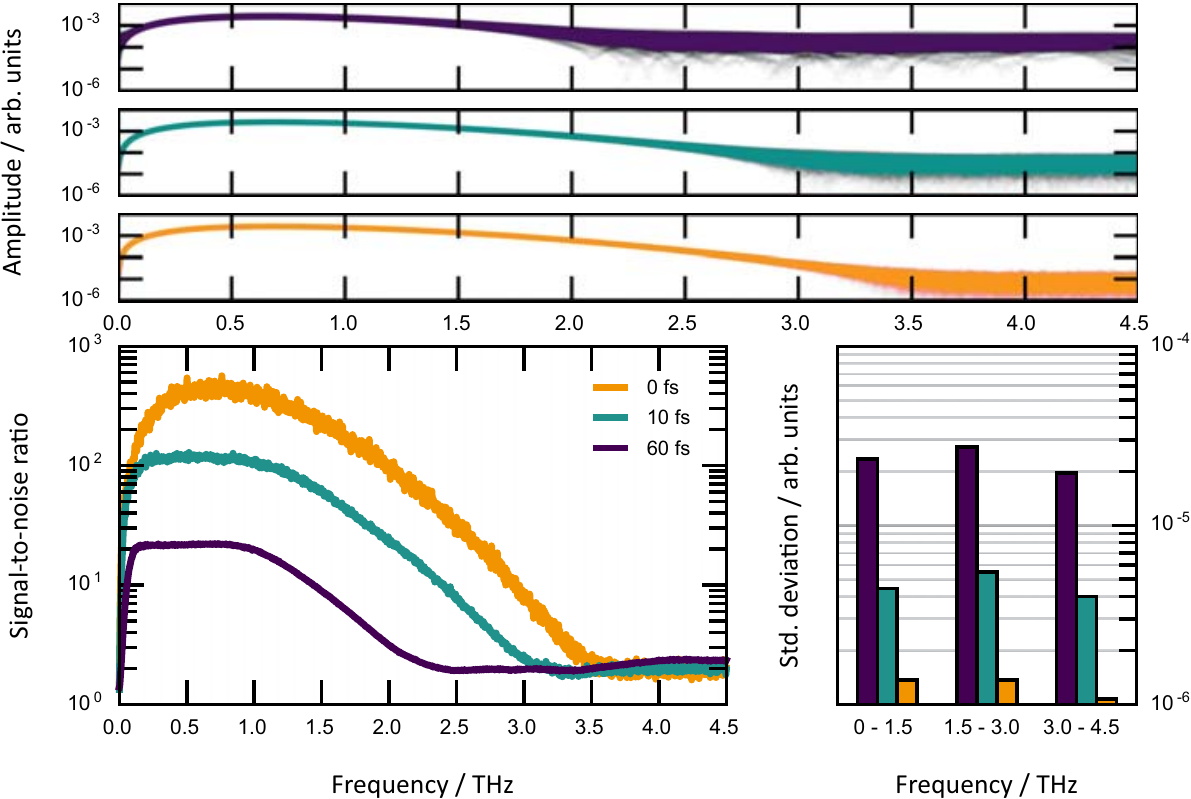}
    \caption{Top: One hundred single shots of simulated dry-air terahertz spectra for three different random time-axis error settings (0 fs, 10 fs, 60 fs). An applied random time-axis error reduces the usable terahertz bandwidth. Left: Signal-to-noise ratio. Jitter reduces the signal-to-noise ratio and thus, the noise floor is reached at a lower frequency already. Right: Standard deviation over frequency. The different jitter levels can be seen clearly.}
    \label{fig:randomSimulation}
\end{figure}

As the simulation shows (Fig.~\ref{fig:randomSimulation}), a random time-axis error translates to a flickering of the FFT-generated spectrum's tail in the frequency domain. One hundred single shots of simulated dry-air terahertz spectra demonstrate this phenomenon for three different jitter settings (0~fs, 10~fs, and 60~fs) (top). The more time-axis jitter is applied, the earlier a tail flickering becomes notable and less usable terahertz bandwidth is available. This effect is further quantified by calculating the signal-to-noise ratio (left) as well as the standard deviation over frequency (right). As expected, an applied time-axis jitter reduces the signal-to-noise ratio and thus the simulated noise floor is already reached at a lower frequency. The standard deviation visualizes the different jitter levels in the frequency domain.

\subsection{Simulation of systematic error}

\begin{figure}[h!]
    \centering\includegraphics[width=0.99\textwidth]{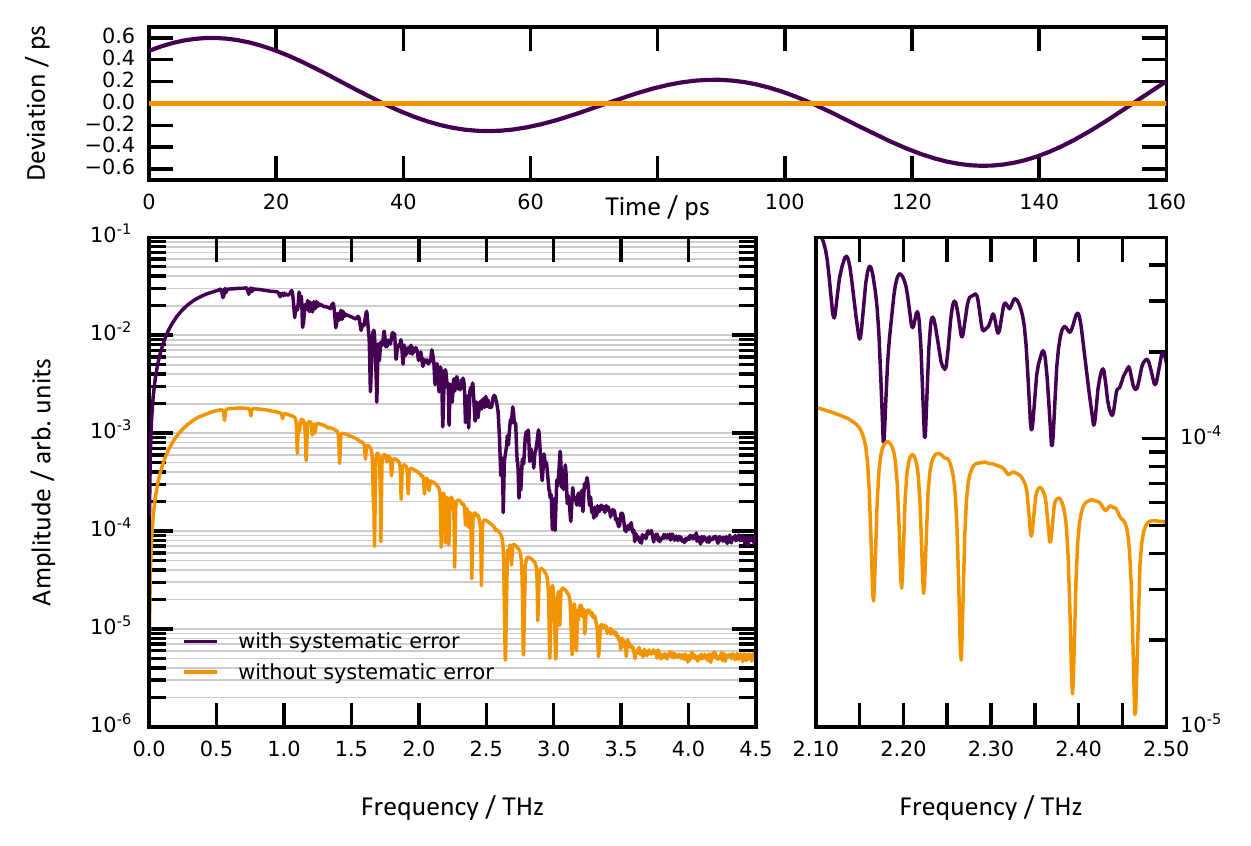}
    \caption{Top: Deviation over phase difference. The simulated systematic error is shown in purple. Bottom: The systematic error distorts the terahertz spectrum which affects the shape and position of the spectral lines. This leads to a pronounced asymmetry and a positional shift. Thus, spectral lines start to interfere and cannot be resolved anymore. The spectra are offset for better visibility.}
    \label{fig:systematicSimulation}
\end{figure}

In contrast to a random error, a systematic error does not reduce the usable bandwidth, but rather distorts the terahertz spectrum. This becomes apparent by taking a close look at the shape and position of the spectral absorption lines (Fig.~\ref{fig:systematicSimulation}). Dependent on the local distortion, the spectral lines develop an asymmetry and are shifted away from their actual position. As a result, spectral lines that are close together interfere and cannot be resolved anymore, rendering the absorption information useless. 

\newpage

\section{Results}

\subsection{Investigation of random error}

\subsubsection{Jitter}

\begin{figure}[h!]
    \centering\includegraphics[width=0.99\textwidth]{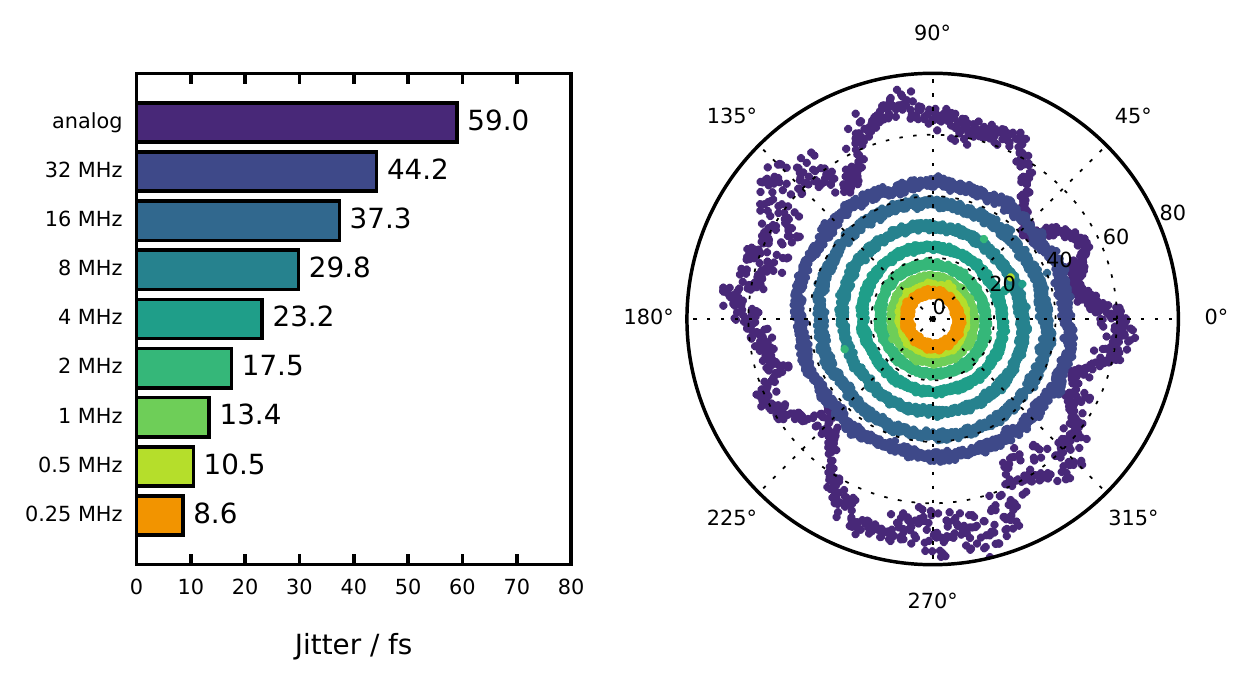}
    \caption{Measured jitter in dependence on phase difference for analog and 16-bit digital phase transmission. The phase-difference dependence of the analog phase transmission is clearly visible (purple). It is mainly caused by a non-constant jitter characteristic of the digital-to-analog converters. A 16-bit digital phase transmission solves this problem resulting in a nearly phase-difference independent jitter. By adjusting the transmission bandwidth, the jitter can be reduced to the sub-10 fs regime (orange). Dependent on the application's requirements the optimal bandwidth-jitter operational point can be chosen freely.}
    \label{fig:jitter}
\end{figure}

The setup shown in Fig.~\ref{fig:setup} practically exhibits four different processes that contribute to the overall random error of the system: optical-to-electronical conversion, analog-to-digital conversion, digital-to-analog conversion, and externally induced noise. We found that the digital-to-analog conversion at the output of the used FPGA causes a significant, non-linear noise contribution which yields a phase-dependent jitter (purple plot in Fig.~\ref{fig:jitter}). To eliminate this error, we switched from an analog phase transmission to a 16-bit digital phase transmission. This did not only remove the digital-to-analog conversion as a non-linear noise source, but also opened the door for transmission-bandwidth control and thus a methodical jitter reduction. Figure~\ref{fig:jitter} demonstrates the results of these achievements. Firstly, the change from an analog to a digital phase transmission eliminated the phase dependence. Secondly, dependent on the applications requirements, a bandwidth-jitter operation point can be chosen freely. For the terahertz experiments presented in this paper, we chose the 250~kHz-8~fs operation point, since the DAQ receives the interferometrically determined phase information at 250 kHz.  

\subsubsection{Terahertz dry-air absorption spectra}

To investigate the influence of the jitter reduction on the terahertz performance, one hundred single shots of dry-air terahertz absorption spectra have been recorded for each phase detection setup (interferometric, analog, and 16-bit digital). Figure~\ref{fig:snr} (top) displays the achieved terahertz performance enhancement. In accordance with the simulation (Fig.~\ref{fig:randomSimulation}), the newly developed digital phase detection setup with sub-10~fs jitter exhibits reduced flickering and increases the practically usable terahertz bandwidth by approx. 1~THz. 

\subsubsection{Signal-To-Noise}

\begin{figure}[h!]
    \centering\includegraphics[width=0.99\textwidth]{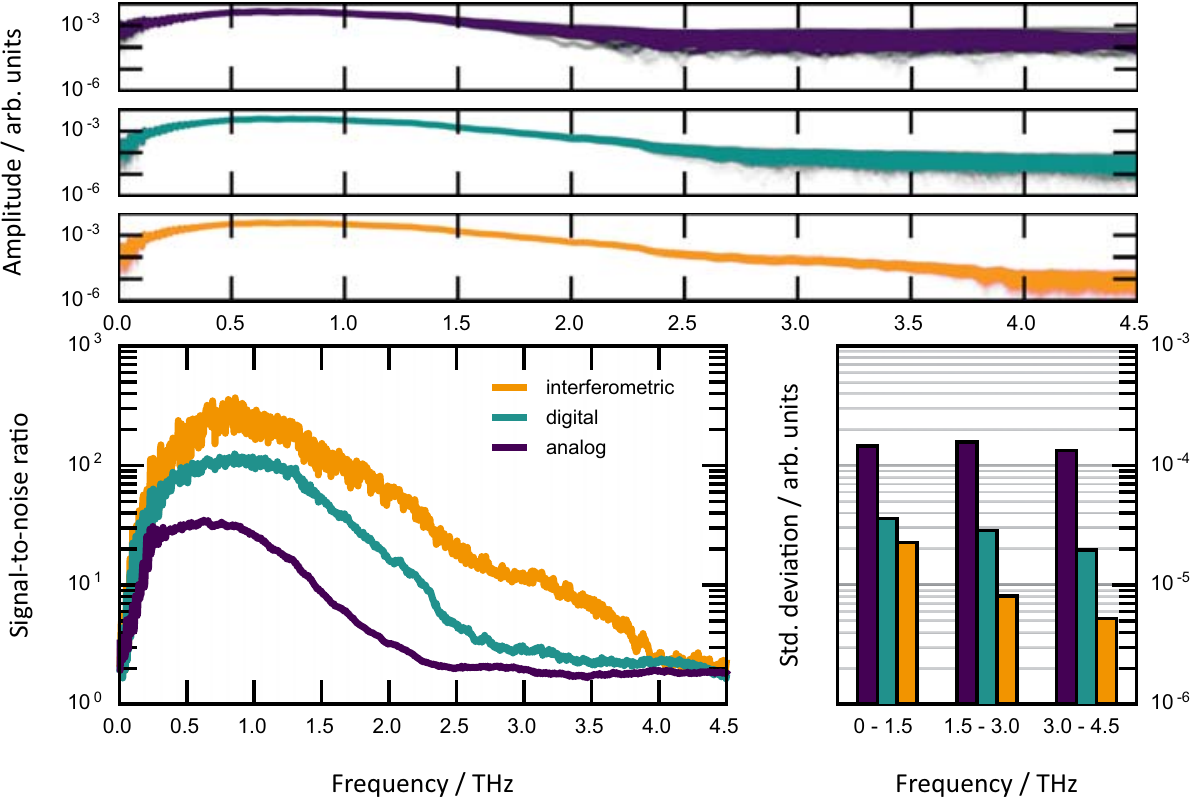}
    \caption{Top: One hundred single shots of dry-air terahertz absorption spectra. The newly developed phase detection unit utilizing a 16-bit digital phase transmission (green) clearly outperforms our previously used phase detection unit utilizing an analog phase transmission (purple): The practically usable terahertz bandwidth is increased by approx. 1~THz. This closes the performance gap to the technically superior interferometric approach (orange) further. Left: signal-to-noise ratio calculated from the recordings above. Compared to the previously used analog transmitting phase detection unit (purple), the signal-to-noise ratio of the newly developed digital transmitting phase detection unit (green) increases by up to a factor of 7, resulting in a practically usable terahertz bandwidth of more than 2.5~THz. The signal-to-noise ratio of the interferometric phase detection unit (orange) illustrates the potential for improvement. Right: standard deviation versus frequency calculated from the same one hundred individual images. The graph confirms the significant improvement from the previous analog to the newly developed digital approach. All results qualitatively agree with the simulation.}
    \label{fig:snr}
\end{figure}

To further investigate the achieved improvement, the signal-to-noise ratio as well as the standard deviation versus frequency were calculated. The results are shown in the bottom half of Fig.~\ref{fig:snr} and are in qualitative agreement with the simulation. Comparing the previously used analog transmitting phase detection unit with the newly developed digital transmitting phase detection unit, an increase of the signal-to-noise ratio by up to a factor of 7 (left side) can be seen. This shifts the practically usable upper terahertz limit to above 2.5~THz, which is sufficient for most technical applications. In addition, the increase in signal-to-noise ratio translates directly into a significant reduction in measurement times. A look at the plot of the standard deviation versus frequency completes the pictures (right side). 

\subsection{Investigation of systematic error}

\subsubsection{Calibration}

Besides a random error causing a time-axis jitter, there is a systematic error causing a static distortion of the time axis. As a result, the spectral absorption lines in the FFT-derived terahertz spectra develop an asymmetry and shift away from their actual position (see simulation). Using the phase difference information acquired by the interferometric phase detection unit as a measurement standard, the systematic error could be determined. The result is shown in Fig.~\ref{fig:distortion}: The systematic error is significant and the absolute value exceeds 600~fs at the largest deviation. However, since this error is static, it can be stored in a look-up table which is then used to calibrate the system. After calibration, the systematic error is removed and only a random error remains.   

\subsubsection{Influence on spectral features}

\begin{figure}[h!]
    \centering\includegraphics[width=0.99\textwidth]{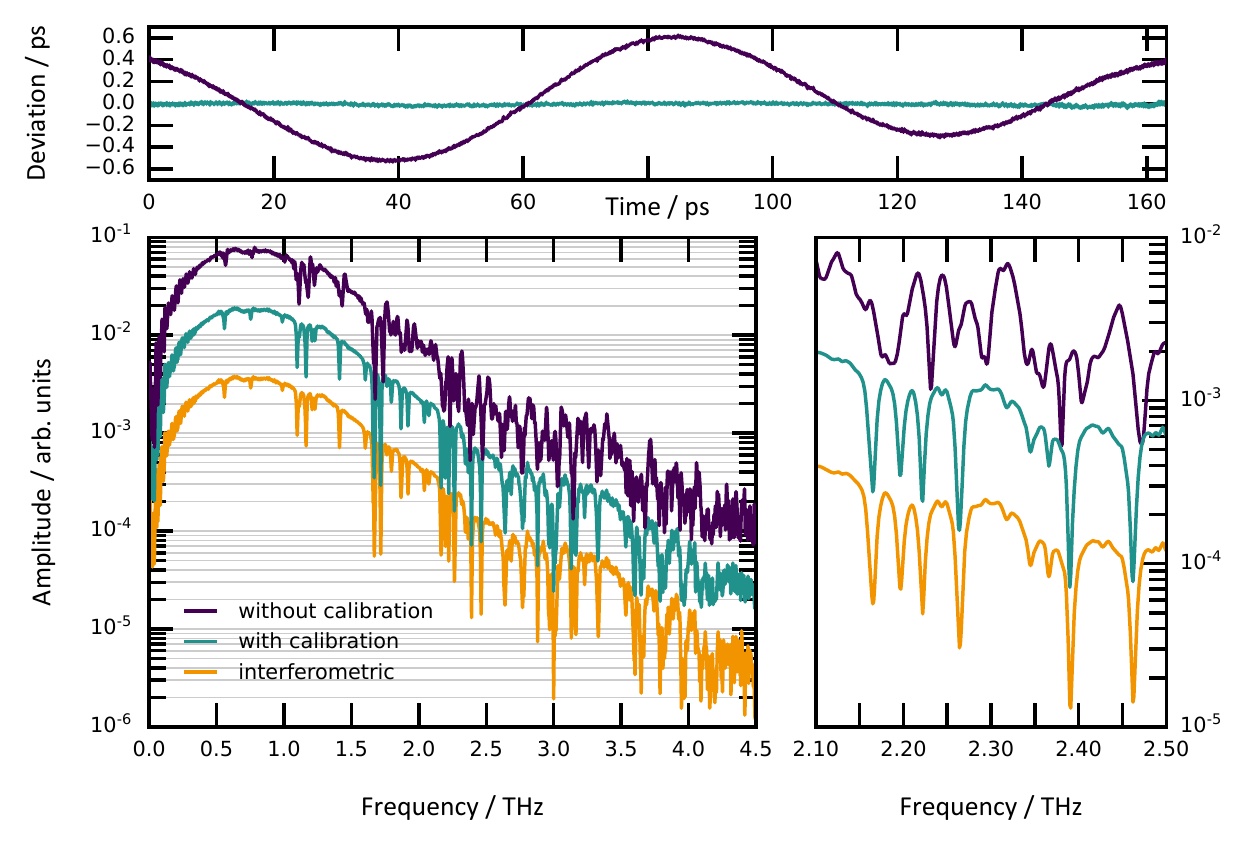}
    \caption{Top: Deviation over phase difference. The systematic error is clearly visible and its absolute value exceeds 600 fs at the largest deviation (purple). Using a look-up-table approach, the system can be calibrated. This results in a nearly systematic-error-free system with only a small random error left (green). Bottom: Mean of 20 ambient-air terahertz absorption spectra recorded with 1~s integration time (offset for better visibility). At first glance, the spectra on the left side look quite similar. However, a closer look reveals the differences in accordance with the simulation: the non-calibrated system is not able to resolve the spectral absorption lines (purple). In contrast, the calibrated system (green) resolves the spectral absorption lines identically to the interferometric measurement standard (orange).}
    \label{fig:distortion}
\end{figure}

The systematic-error-induced distortion of the terahertz absorption spectra becomes evident by taking a close look at the spectral absorption lines of water vapor in air. Figure~\ref{fig:distortion} shows the mean of 20 ambient-air terahertz absorption spectra, each recorded with an integration time of 1~s, for the newly developed digital transmitting phase detection unit without calibration (purple) and with calibration (green) as well as for the interferometric measurement standard (orange). The zoom on the right side exemplarily reveals the improvement achieved by the calibration: contrary to the non-calibrated system, the calibrated system is able to resolve the spectral absorption lines and performs as good as the measurement standard, nearly without any noticeable difference. 

\subsection{Multi-layer thickness measurement}

\begin{figure}[h!]
    \centering\includegraphics[width=0.99\textwidth]{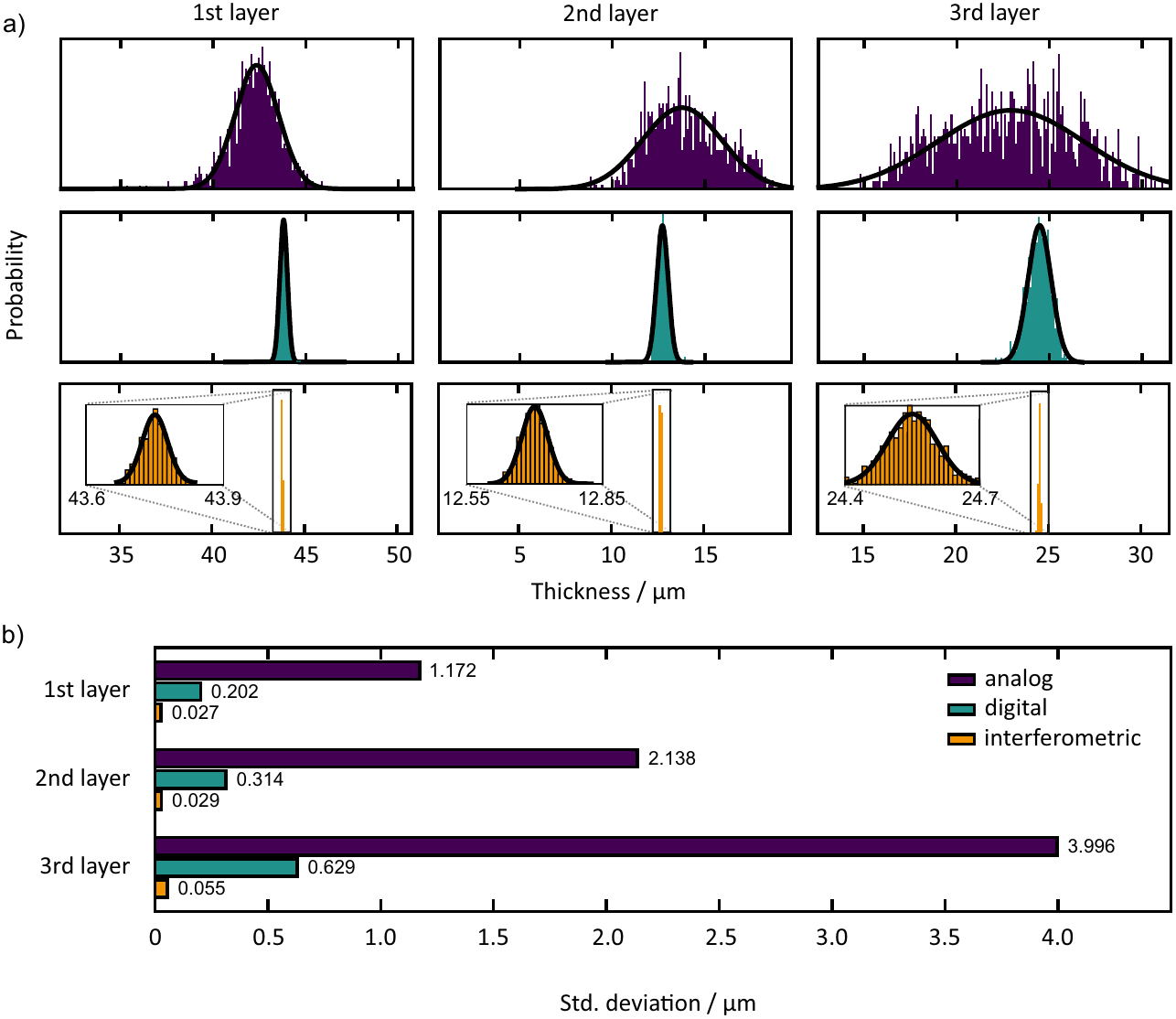}
    \caption{a) Layer-thickness measurement results of a three-layer coating on metal of 1000 consecutive measurements each. The newly developed digital transmitting phase-detection unit outperforms the previously used unit clearly (green vs. purple). The performance of the interferometric measurement standard reveals the potential for further innovations (orange). b) Standard deviation for the three phase-detection systems under test. The standard deviation of the electronic phase-detection unit could be reduced by a factor of 5.8, 6.8, and 6.4 for the first, second, and third layer, respectively, pushing the overall performance well into the sub-µm regime (green vs. purple). }
    \label{fig:thicknessMeasurement}
\end{figure}

Non-destructive, contactless, multi-layer thickness measurement is an established industrial terahertz application. By knowing the refractive indices of the layers under test, the thickness of each layer can be derived from extended analysis of the reflected (or transmitted) terahertz signal even for sub-wavelength thicknesses~\cite{Ellrich2020, Krimi2016, Pfeiffer2018}. The higher the resolution of the recorded terahertz time-domain signal, the higher the accuracy of the layer thickness determination. Figure~\ref{fig:thicknessMeasurement}~a) exemplarily visualizes the results of 1000 consecutive measurements for a three-layer system dependent on the used phase detection unit. Clearly, the reduced timing jitter and systematic error of the newly developed digital transmitting electronic phase detection unit yields a convincing accuracy enhancement in comparison to the previously used analog transmitting electronic phase detection unit. Similar to the signal-to-noise ratios displayed in  Fig.~\ref{fig:snr}, meeting the performance of the interferometric phase detection approach is still a few innovations ahead. However, as shown in Fig.~\ref{fig:thicknessMeasurement}~b), the standard deviation could be pushed well into the sub-µm regime which is fairly sufficient for most industrial applications.

\newpage

\section{Conclusions}

We have presented the recent design wins of our electronic phase detection unit development which include a more than six-fold reduction of timing jitter reaching the sub-10~fs regime, as well as a calibration technique to remove almost any systematic error. We achieved this by switching from an analog I-Q phase transmission to a 16-bit digital phase transmission which allows for application-specific transmission bandwidth control, and by utilizing our interferometrically monitored delay line with an accuracy of about 1.1~fs as a calibration standard. This led to a significant increase in exploitable terahertz bandwidth and signal-to-noise ratio, as well as an distortion-free resolution of spectral absorption lines. These results were substantiated by an extensive simulation which is in excellent qualitative agreement with the measurements. Finally, we demonstrated that our design wins yield a more than five-fold accuracy improvement of our electronic-phase-detection-based multi-layer thickness measurements. These improvements represent an important step towards higher resolution at faster measurement times. 

\newpage

\end{document}